\def\year{2020}\relax
\documentclass[letterpaper]{article} 
\usepackage{aaai20}  
\usepackage{times}  
\usepackage{helvet} 
\usepackage{courier}  
\usepackage[hyphens]{url}  
\usepackage{graphicx} 
\usepackage{graphicx}  
\usepackage{amsmath}
\usepackage{amssymb}
\usepackage{amsfonts}
\usepackage{amsthm}
\usepackage{amstext}
\usepackage{mathtools}
\usepackage{afterpage}
\usepackage{capt-of}

\usepackage{caption}
\usepackage{subcaption}

\usepackage[utf8]{inputenc}
\usepackage{breakcites}
\usepackage[square,numbers]{natbib}
\usepackage{enumitem}
\usepackage{appendix}
\usepackage{booktabs}
\usepackage{array}
\usepackage{colortbl}
\usepackage{longtable}

\usepackage{algorithm}
\usepackage[noend]{algpseudocode}

\usepackage{multirow}
\usepackage{multicol}

\usepackage{todonotes}

\title{Automated Database Indexing Using Model-Free Reinforcement Learning}

\author{
Gabriel Paludo Licks$^{\dag}$ and \textbf{Felipe Meneguzzi\ddag} \\
Pontifical Catholic University of Rio Grande do Sul (PUCRS), Brazil \\
Graduate Program in Computer Science, School of Technology \\
$^{\dag}$\texttt{gabriel.licks@edu.pucrs.br} \\
\ddag\texttt{felipe.meneguzzi@pucrs.br} \\
}

\begin{document}

\maketitle

\begin{abstract}
Configuring databases for efficient querying is a complex task, often carried out by a database administrator. 
Solving the problem of building indexes that truly optimize database access requires a substantial amount of database and domain knowledge, the lack of which often results in wasted space and memory for irrelevant indexes, possibly jeopardizing database performance for querying and certainly degrading performance for updating. 
We develop an architecture to solve the problem of automatically indexing a database by using reinforcement learning to optimize queries by indexing data throughout the lifetime of a database. 
In our experimental evaluation, our architecture shows superior performance compared to related work on reinforcement learning and genetic algorithms, maintaining near-optimal index configurations and efficiently scaling to large databases. 
\end{abstract}

\section{Introduction}

Despite the multitude of tools available to manage and gain insights from very large datasets, indexing databases that store such data remains a challenge with multiple opportunities for improvement~\cite{wang2016learning}. 
Slow information retrieval in databases entails not only wasted time for a business but also indicates a high computational cost being paid. 
Unnecessary indexes or columns that should be indexed but are not,
directly impact the query performance of a database. 
Nevertheless, achieving the best indexing configuration for a database is not a trivial task~\cite{duan2009tuning, elfayoumy2012database}. 
To do so, we have to learn from queries that are running, take into account their performance, the system resources, and the storage budget so that we can find the best index candidates~\cite{petraki2015holistic}. 


In an ideal scenario, all frequently queried columns should be indexed to optimize query performance. 
Since creating and maintaining indexes incur a cost in terms of storage as well as in computation whenever database insertions or updates take place in indexed columns~\cite{ramakrishnan2000database}, choosing an optimal set of indexes for querying purposes is not enough to ensure optimal performance, so we must reach a trade-off between query and insert/update performance.
Thus, this is a fundamental task that needs to be performed continuously, as the indexing configuration directly impacts on a database's overall performance.

We developed an architecture for automated and dynamic database indexing that evaluates query and insert/update performance to make decisions on whether to create or drop indexes using Reinforcement Learning (RL). 
We performed experiments using a scalable benchmark database, where we empirically evaluate our architecture results in comparison to standard baseline index configurations, database advisor tools, genetic algorithms, and other reinforcement learning methods applied to database indexing.
The architecture we implemented to automatically manage indexes through reinforcement learning successfully converged in its training to a configuration that outperforms all baselines and related work, both in performance and in storage usage by indexes.

\section{\label{sec:background}Background}


\subsection{\label{sec:rl}Reinforcement Learning}


Reinforcement learning is an approach to learn optimal agent policies in stochastic environments modeled as Markov Decision Processes (MDPs)~\cite{bellman1957markovian}. 
It is characterized by a trial-and-error learning method, where an agent interacts and transitions through states of an MDP environment model by taking actions and observing rewards~\cite[Ch. 1]{sutton2018reinforcement}. 
MDP are formally defined as a tuple $\mathcal{M} = \langle \mathcal{S}, \mathcal{A}, \mathcal{P}, \mathcal{R}, \gamma \rangle$, where $\mathcal{S}$ is the state space, $\mathcal{A}$ is the action space, $\mathcal{P}$ is a transition probability function which defines the dynamics of the MDP, $\mathcal{R}$ is a reward function and $\gamma \in [0, 1]$ is a discount factor~\cite[Ch. 3]{sutton2018reinforcement}.

In order to solve an MDP, an agent needs to know the state-transition and the reward functions. 
However, in most realistic applications, modeling knowledge about the state-transition or the reward function is either impossible or impractical, so an agent interacts with the environment taking sequential actions to collect information and explore the search space by trial and error~\cite[Ch. 1]{sutton2018reinforcement}.
The Q-learning algorithm is the natural choice for solving such MDPs~\cite[Ch. 16]{sutton2018reinforcement}.
This method learns the values of state-action pairs, denoted by $Q(s, a)$, representing the value of taking action $a$ in a state $s$~\cite[Ch. 6]{sutton2018reinforcement}.
%
%

Assuming that states can be described in terms of features that are well informative, such problem can be handled by using linear function approximation, which is to use a parameterized representation for the state-action value function other than a look-up table~\cite{tsitsiklis1996analysis}. 
The simplest differentiable function approximator is through a linear combination of features, though there are other ways of approximating functions such as using neural networks~\cite[Ch. 9, p. 195]{sutton2018reinforcement}.

\subsection{\label{sec:dbIndexing}Indexing in Relational Databases}

An important technique to file organization in a DBMS is \textit{indexing}~\cite[Ch. 8, p. 274]{ramakrishnan2000database}, and is usually managed by a DBA.
However, index selection without the need of a domain expert is a long-time research subject and remains a challenge due to the problem complexity~\cite{wang2016learning, elfayoumy2012database, duan2009tuning}.
The idea is that, given the database schema and the workload it receives, we can define the problem of finding an efficient index configuration that optimizes database operations~\cite[Ch. 20, p. 664]{ramakrishnan2000database}.
The complexity stems from the potential number of attributes that can be indexed and all of its subsets.
%
%

While DBMSs strive to provide automatic index tuning, the usual scenario is that performance statistics for optimizing queries and index recommendations are offered, but the DBA makes the decision on whether to apply the changes or not.
Most recent versions of DBMSs such as Oracle~\cite{olofson2018ensuring} and Azure SQL Database~\cite{Azure} can automatically adjust indexes.
However, it is not the case that the underlying system is openly described.

A general way of evaluating DBMS performance is through benchmarking.
Since DBMSs are complex pieces of software, and each has its own techniques for optimization, external organizations have defined protocols to evaluate their performance~\cite[Ch. 20, p. 682]{ramakrishnan2000database}.
The goals of benchmarks are to provide measures that are portable to different DBMSs and evaluate a wider range of aspects of the system, e.g., transactions per second and price-performance ratio~\cite[Ch. 20, p. 683]{ramakrishnan2000database}.

\subsection{\label{sec:tpchBenchmark}TPC-H Benchmark}


The tools provided by TPC-H include a database generator (DBGen) able to create up to 100 TB of data to load in a DBMS, and a query generator (QGen) that creates 22 queries with different levels of complexity. 
Using the database and workload generated using these tools, TPC-H specifies a benchmark that consists of inserting records, executing queries, and deleting records in the database to measure the performance of these operations. 

The TPC-H Performance metric is expressed in Queries-per-Hour ($QphH@Size$), which is achieved by computing the $Power@Size$ and the $Throughput@Size$ metrics~\cite{thanopoulou2012benchmarking}. 
The resulting values are related to its scale factor ($@Size$), i.e., the database size in gigabytes. 
The $Power@Size$ evaluates how fast the DBMS computes the answers to single queries. 
This metric is computed using Equation~\ref{eq:1}:

\begin{equation}
\small
\label{eq:1}
Power@Size = \frac{3600} {\sqrt[24]{\pi_{i=1}^{22} QI(i, 0) \times \pi_{j=1}^{2} RI(j, 0)}} \times SF
\end{equation}

\noindent where $3600$ is the number of seconds per hour and $QI(i, s)$ is the execution time for each one of the queries $i$. $RI(j, s)$ is the execution time of refresh functions $j$ (insert/update) in the query stream $s$, and $SF$ is the scale factor or database size, ranging from $1$ to $100,000$ according to its $@Size$. 

The $Throughput@Size$ measures the ability of the system to process the most queries in the least amount of time, taking advantage of I/O and CPU parallelism~\cite{thanopoulou2012benchmarking}. It computes the performance of the system against a multi-user workload performed in an elapsed time, using Equation~\ref{eq:2}:

\begin{equation}
\small
\label{eq:2}
Throughput@Size = \frac{S \times 22} {T_{S}^{}} \times 3600 \times SF
\end{equation}

\noindent where $S$ is the number of query streams executed, and $T_{S}$ is the total time required to run the test for $s$ streams. 

\begin{equation}
\small
\label{eq:3}
QphH@Size = \sqrt {Power@Size \times Throughput@Size}
\end{equation}

Equation~\ref{eq:3} shows the Query-per-Hour Performance ($QphH@Size$) metric, which is obtained from the geometric mean of the previous two metrics and reflects multiple aspects of the capability of a database to process queries. 
The $QphH@Size$ metric is the final output metric of the benchmark and summarizes both single-user and multiple-user overall database performance.

\section{\label{sec:architecture}Architecture}

In this section, we introduce our database indexing architecture to automatically choose indexes in relational databases, which we refer to as SmartIX.
The main motivation of SmartIX is to abstract the database administrator's task that involves a frequent analysis of all candidate columns and verifying which ones are likely to improve the database index configuration. 
For this purpose, we use reinforcement learning to explore the space of possible index configurations in the database, aiming to find an optimal strategy over a long time horizon while improving the performance of an agent in the environment. 

\begin{figure}[b]
	\centering
	\includegraphics[width=0.3\textwidth]{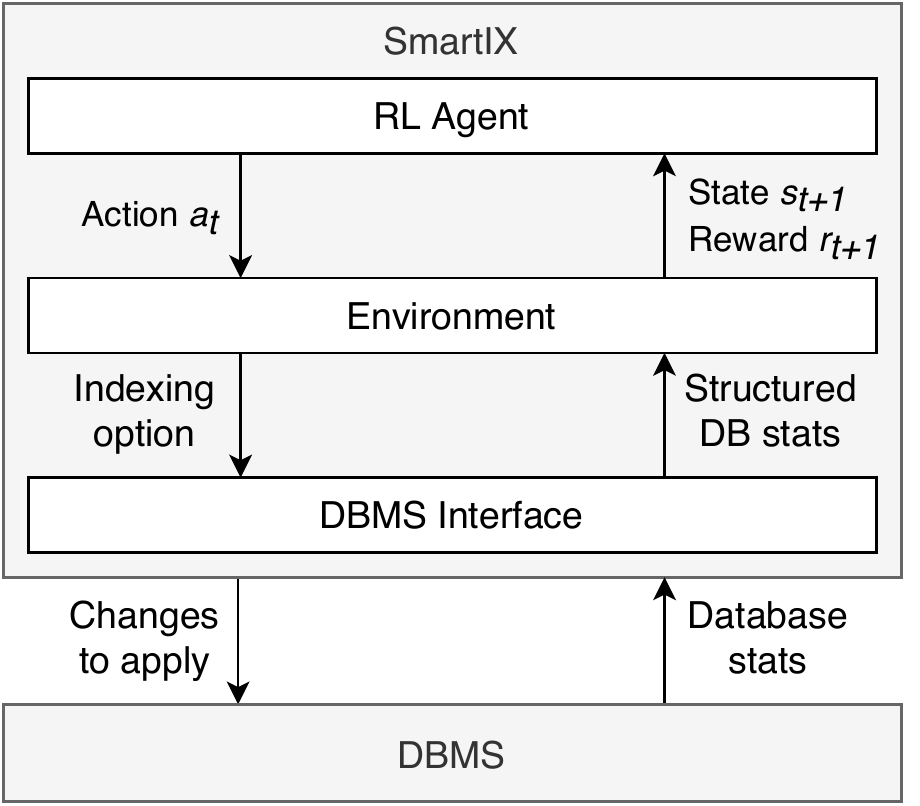}
	\caption{SmartIX architecture.}
	\label{fig:agent_state_actions}
\end{figure}

The SmartIX architecture is composed of a reinforcement learning agent, an environment model of a database, and a DBMS interface to apply agent actions to the database. 
The reinforcement learning agent is responsible for the decision making process.
The agent interacts with an environment model of the database, which computes system transitions and rewards that the agent receives for its decisions.
To make changes persistent, there is a DBMS interface that is responsible for communicating with the DBMS to create or drop indexes and get statistics of the current index configuration.

\subsection{\label{sec:agent}Agent}

Our agent is based on the Deep Q-Network agent proposed by~\cite{mnih2015human}.
Algorithm~\ref{alg:agent-algorithm} consists of a reinforcement learning method built around the Q-learning, using a neural network for function approximation, and a replay memory for experience replay. 
The neural network is used to approximate the action-value function and is trained using mini-batches of experience randomly sampled from the replay memory.
At each time step, the agent performs one transition in the environment.
That is, the agent chooses an action using an epsilon-greedy exploration function at the current state, the action is then applied in the environment, and the environment returns a reward signal and the next state.
Finally, each transition in the environment is stored in the replay buffer, and the agent performs a mini-batch update in the action-value function.

\begin{algorithm}
    \caption{Indexing agent with function approximation and experience replay. From~\cite[Ch. 6, p. 131]{sutton2018reinforcement} and~\cite{mnih2015human}.}
    \label{alg:agent-algorithm}
    \begin{algorithmic}[1]
        \small
        \State Random initialization of the value function
        \State Empty initialization of a replay memory $D$
        \State $s \gets DB\;initial\;index\;configuration$
        \For{each step}
            \State $a \gets epsilon\:greedy(s)$
            \State $s', r \gets execute(a)$
            \State Store experience $e = \langle s, a, r, s' \rangle$ in $D$
            \State Sample random mini-batch of experiences $e \sim D$
            \State Perform experience replay using mini-batch
            \State{$s \gets s'$}
        \EndFor
    \end{algorithmic}
\end{algorithm}

\subsection{Environment}

The environment component is responsible for computing transitions in the system and computing the reward function. 
To successfully apply a transition, we implement a model of the database environment, modeling states that contain features that are relevant to the agent learning, and a transition function that is able to modify the state with regard to the action an agent chooses.
Each transition in the environment outputs a reward signal that is fed back to the agent along with the next state, and the reward function has to be informative enough so that the agent learns which actions yield better decisions at each state.

\subsubsection{\label{sec:agent-state}State representation}

The state is the formal representation of the environment information used by the agent in the learning process. 
Thus, deciding which information should be used to define a state of the environment is critical for task performance. 
The amount of information encoded in a state imposes a trade-off for reinforcement learning agents.
Specifically, that if the state encodes too little information, then the agent might not learn a useful policy, whereas if the state encodes too much information, there is a risk that the learning algorithm needs too many samples of the environment that it does not converge to a policy. 

For the database indexing problem, the state representation is defined as a feature vector $\vec{S} = \vec{I}\cdot\vec{Q}$, which is a result of a concatenation of the feature vectors $\vec{I}$ and $\vec{Q}$.
The feature vector $\vec{I}$ encodes information regarding the current index configuration of the database, with length $|\vec{I}| = C$, where $C$ is a constant of the total number of columns in the database schema. 
Each element in the feature vector $\vec{I}$ holds a binary value, containing $1$ or $0$, depending on whether the column that corresponds to that position in the vector is indexed or not. 
The second part of our state representation is a feature vector $\vec{Q}$, also with length $|\vec{Q}| = C$, which encodes information regarding which indexes were used in last queries received by the database.
To organize such information, we set a constant value of $H$ that defines the horizon of queries that we keep track of. 
To each of the last queries in a horizon $H$, we verify whether any of the indexes currently created in the database are used to run such queries. 
Each position in the vector $\vec{Q}$ corresponds to a column and holds a binary value that is assigned $1$ if such column is indexed and used in the last $H$ queries, else $0$.
Finally, the concatenate both $\vec{I}$ and $\vec{Q}$ to generate our binary state vector $\vec{S}$ with length $|\vec{S}| = 2C$.

\subsubsection{Actions}

In our environment, we define the possible actions as a set $A$ of size $C+1$.
Each one of the $C$ actions refers to one column in the database schema.
These actions are implemented as a ``flip'' to create or drop an index in the current column.
Therefore, for each action, there are two possible behaviors: \textsc{create index} or \textsc{drop index} on the corresponding column.
The last action is a ``do nothing'' action, that enables the agent not to modify the index configuration in case it is not necessary at the current state.


\subsubsection{Reward}

Deciding the reward function is critical for the quality of the ensuing learned policy. 
On the one hand, we want the agent to learn that indexes that are used by the queries in the workload must be maintained in order to optimize such queries.
On the other hand, indexes that are not being used by queries must not be maintained as they consume system resources and are not useful to the current workload.
Therefore, we compute the reward signal based on the next state's feature vector $\vec{S}$ after an action is applied, since our state representation encodes information both on the current index configuration and on the indexes used in the last queries, i.e. information contained in vectors $\vec{I}$ and $\vec{Q}$.
Our reward function is computed using Equation~\ref{eq:reward-function}:

\begin{equation}
\small
\label{eq:reward-function}
\begin{split}
R(op, use) = (1 - op)((1 - use)(1) + (use)(-5)) \\
+ (op)((1-use)(-5) + (use)(1))
\end{split}
\end{equation}

where $op = I_{c}$ and $use = Q_{c}$. 
That is, the first parameter $op$ holds $0$ if the last action represents a dropped index in column $c$, or $1$ if created an index.
The latter parameter, $use$, holds $0$ if an index in column $c$ is not being used by the last $H$ horizon queries, and $1$ otherwise.

Therefore, our reward function returns a value of $+1$ if an index is created and it actually benefits the current workload, or if an index is dropped and it is not beneficial to the current workload. 
Otherwise, the function returns $-5$ to penalize the agent if an index is dropped and it is beneficial to the current workload, or an index is created and it does not benefit the current workload.
The choice of values $+1$ and $-5$ is empirical.
However, we want the penalization value to be at least twice smaller than the $+1$ value, so that the values do not get canceled when accumulating with each other. 
Finally, if the action corresponds to a ``do nothing'' operation, the environment simply returns a reward of $0$, without computing Equation~\ref{eq:reward-function}.

\section{\label{sec:experiments}Experiments}

\subsection{\label{sec:experimental-setup}Experimental setup}


\subsubsection{\label{sec:database-setup}Database setup}

For experimental purposes and due to its usage in literature for measuring database performance, we choose to run experiments using the database schema and data provided by the TPC-H benchmark.
The tools provided by TPC-H include a data generator (DBGen), which is able to create up to 100TB of data to load in a DBMS, and a query generator (QGen) that creates 22 queries with different levels of complexity.
The database we use in these experiments is populated with 1GB of data.
To run benchmarks using each baseline index configuration, we implemented the TPC-H benchmark protocol using a Python script that runs queries, fetches execution time, and computes the performance metrics.

To provide statistics on the database, we show the in Table~\ref{tab:tpch-schema} the number of columns that each table contains and an analysis on the indexing possibilities. 
For that, we mapped for each table in the TPC-H database the total number of columns, the columns that are already indexed (primary and foreign keys, indexed by default), and the remaining columns that are available for indexing. 

\begin{table}
\centering
\caption{TPC-H database - Table stats and indexes}
\vspace{-0.2cm}
\small
\label{tab:tpch-schema}
\begin{tabular}{llccc}
\textbf{Table} & \textbf{Total} & \textbf{Indexed} & \textbf{Indexable} \\ 
\toprule 
\textsc{region} & 3 & 1 & 2 \\
\textsc{nation} & 4 & 2 & 2 \\
\textsc{part} & 9 & 1 & 8 \\
\textsc{supplier} & 7 & 2 & 5 \\
\textsc{partsupp} & 5 & 2 & 3 \\
\textsc{customer} & 8 & 2 & 6 \\
\textsc{orders} & 9 & 2 & 7 \\
\textsc{lineitem} & 16 & 4 & 12 \\
\midrule 
\textbf{Totals} & 61 & 16 & 45 
\end{tabular}
\end{table}

By summing the number of indexable columns in each table, we have a total of 45 columns that are available for indexing. 
Since a column is either indexed or not, there are two possibilities for each of the remaining 45 indexable columns. 
This scenario indicates that we have exactly $35,184,372,088,832$ ($2^{45}$), i.e. more than 35 trillion, possible configurations of \textit{simple} indexes. 
Thus, this is also the number of states that can be assumed by the database indexing configuration and therefore explored by the algorithms.

For comparison purposes, we manually check which columns compose the ground truth optimal index configuration.
We manually create each index possibility and check whether an index benefits at least one query within the 22 TPC-H queries. 
To check whether an index is used or not, we run the \textsc{explain} command to visualize the execution plan of each query.
Finally, we have 6 columns from the TPC-H that compose our ground truth optimal indexes: \textsc{c\_acctbal}, \textsc{l\_shipdate}, \textsc{o\_orderdate}, \textsc{p\_brand}, \textsc{p\_container}, \textsc{p\_size}.

\subsubsection{\label{sec:baselines}Baselines}

The baselines comprise different indexing configurations using different indexing approaches, including commercial and open-source database advisors, and related work on genetic algorithms and reinforcement learning methods. 
Each baseline index configuration is a result of training or analyzing the same workload of queries, from the TPC-H benchmark, in order to make an even comparison between the approaches.
The following list briefly introduces each of them. \textit{Default}: indexes only on primary and foreign keys; \textit{All indexed}: all columns indexed. \textit{Random}: indexes randomly explored by an agent; \textit{EDB}~\citeyear{EDB} and \textit{POWA}~\citeyear{POWA}: indexes obtained using a comercial and an open-source advisor tool, respectively. \textit{ITLCS}~\citeyear{pedrozo2018adaptive} and \textit{GADIS}~\citeyear{neuhausgadis}: indexes obtained using genetic algorithms related work; \textit{NoDBA}~\citeyear{sharma2018case} and \textit{rCOREIL}~\citeyear{basu2016regularized}: indexes obtained using reinforcement learning related work.

The EDB~\citeyear{EDB}, POWA~\citeyear{POWA}, and ITLCS~\citeyear{pedrozo2018adaptive} index configurations are a result of a study conducted by Pedrozo, Nievola and Ribeiro \citeyear{pedrozo2018adaptive}.
The authors~\citeyear{pedrozo2018adaptive} employ these methods to verify which indexes are suggested by each method to each of the 22 queries in the TPC-H workload, whose indexes constitute the respective configurations we use in this analysis.
The index configurations of GADIS~\citeyear{neuhausgadis}, NoDBA~\citeyear{sharma2018case}, and rCOREIL~\citeyear{basu2016regularized} are a result of experiments we ran using source-code provided by the authors. 
We execute the author's algorithms without modifying any hyper-parameter except configuring the database connection.
The index configuration we use in this analysis is the one in which each algorithm converged to, when the algorithm stops modifying the index configuration or reaches the end of training.

\begin{figure*}[tb]
    \centering
    \begin{subfigure}[t]{0.25\textwidth}
        \centering
        \includegraphics[width=\textwidth]{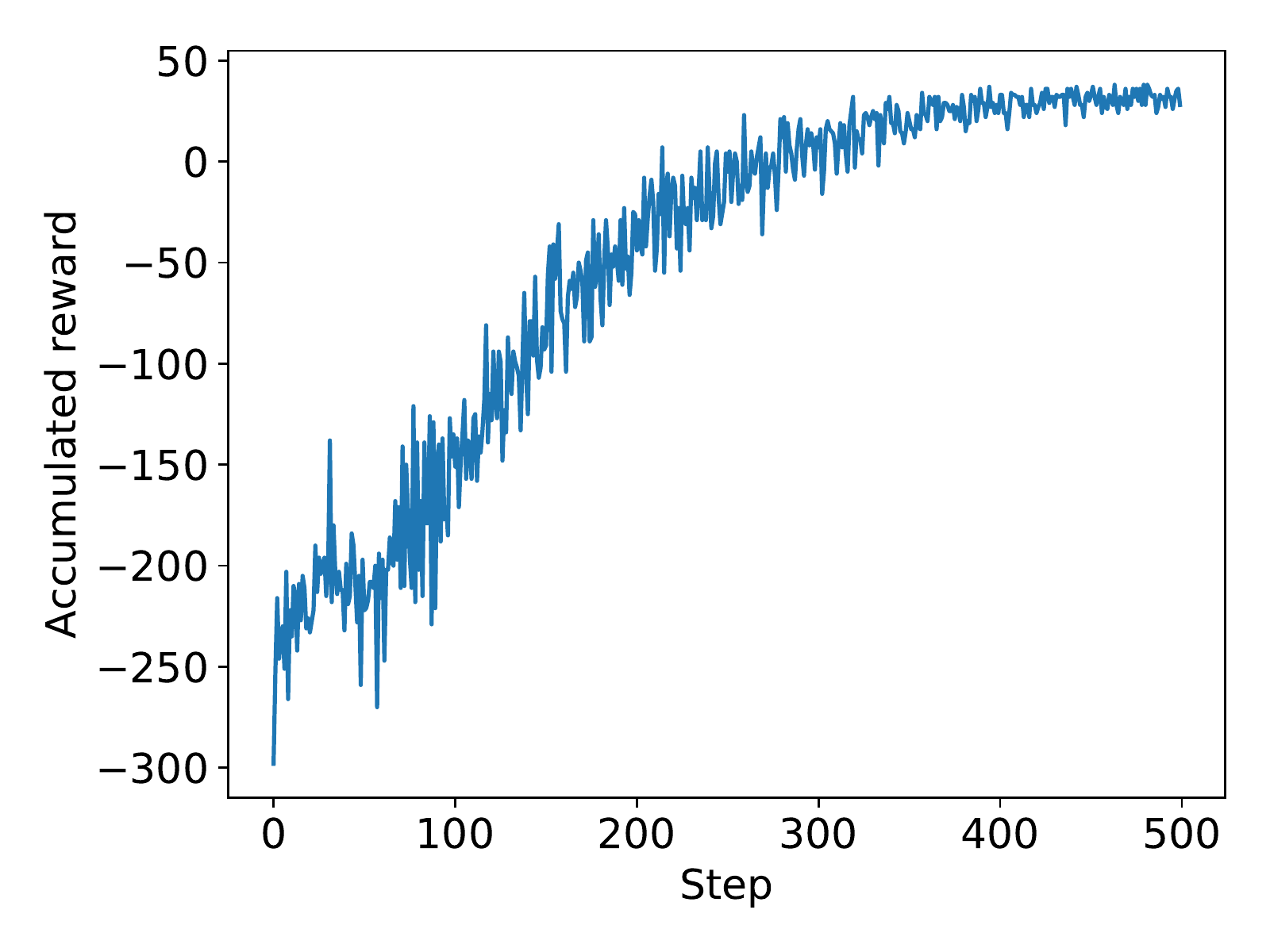}
        \vspace{-0.6cm}
        \caption{Accumulated reward per 128 steps.}
        \label{fig:rewards}
    \end{subfigure}
    \hspace{0.5cm}
    \begin{subfigure}[t]{0.25\textwidth}
        \centering
        \includegraphics[width=\textwidth]{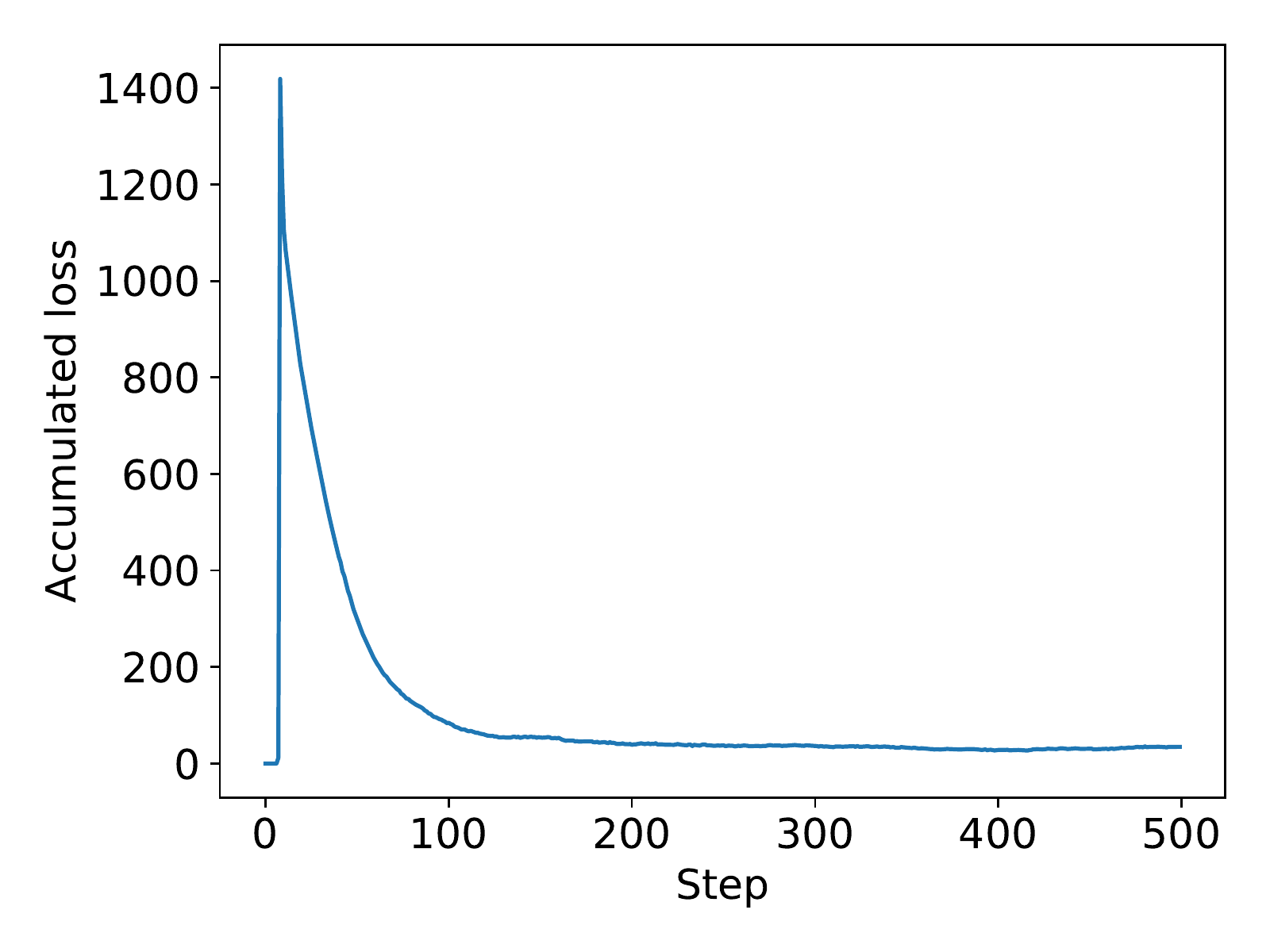}
        \vspace{-0.6cm}
        \caption{Accumulated loss per 128 steps.}
        \label{fig:loss}
    \end{subfigure}
    \hspace{0.5cm}
    \begin{subfigure}[t]{0.25\textwidth}
        \centering
        \includegraphics[width=\textwidth]{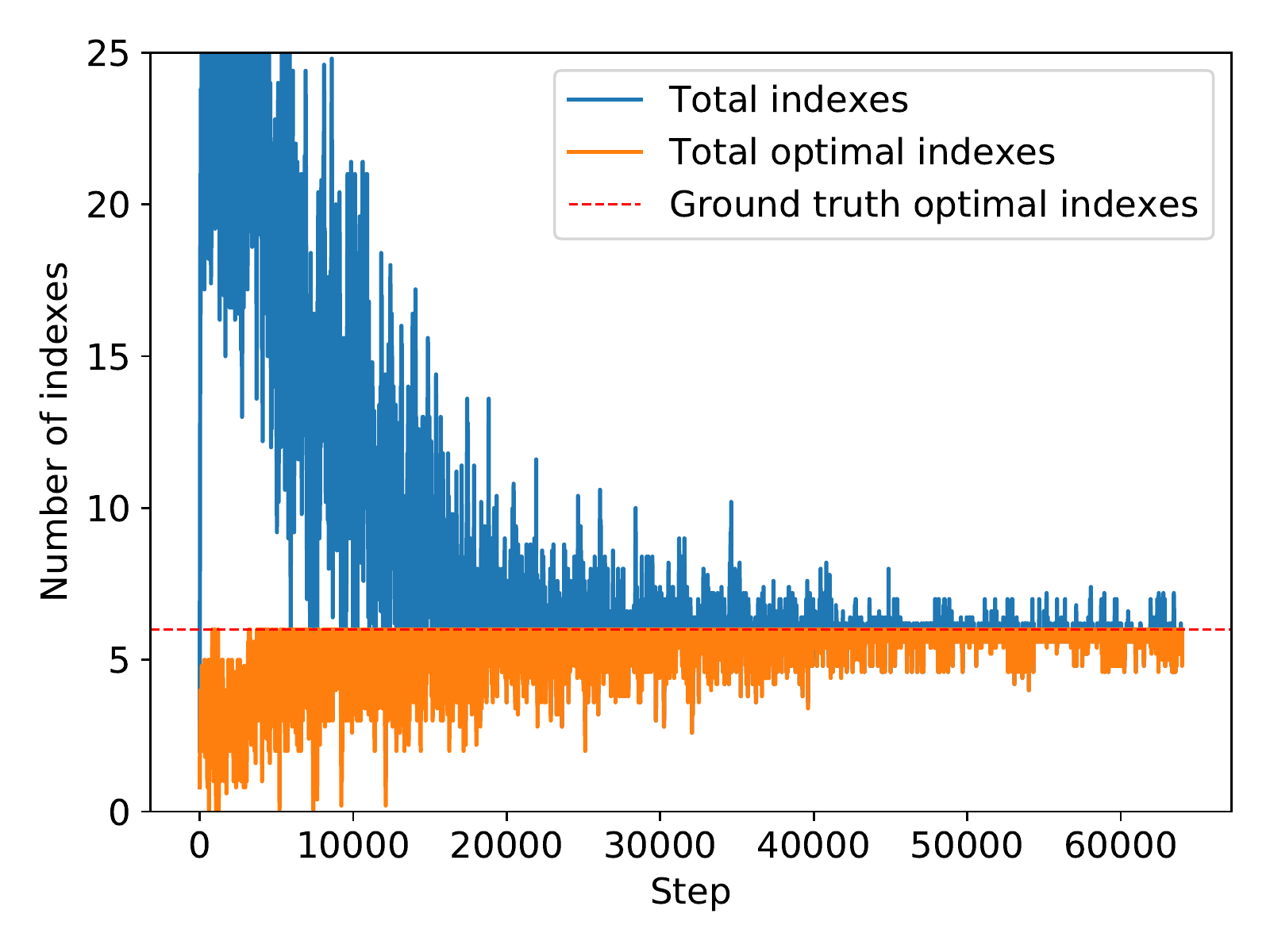}
        \vspace{-0.6cm}
        \caption{Index configurations while training.}
        \label{fig:training-indexes}
    \end{subfigure}
    \vspace{-0.1cm}
    \caption{Training statistics.}
    \label{fig:training-statistics}
\end{figure*}

\subsection{\label{sec:training}Agent training}

Training the reinforcement learning agent consists of time steps of agent-environment interaction and value function updates until it converges to a policy as desired.
In our case, to approximate the value function, we use a simple multi-layer perceptron neural network with two hidden layers and ReLU activation, and an Adam optimizer with mean-squared error loss, both PyTorch 1.5.1 implementations using default hyperparameters~\cite{NEURIPS2019_9015}.
The input and output dimensions depend on the number of columns available to index in the database schema, as shown in Section~\ref{sec:agent-state}.

The hyperparameters used while training are set as follows. 
The first, \textit{learning rate} $\alpha = 0.0001$ and \textit{discount factor} $\gamma = 0.9$, are used in the update equation of the value function.
The next are related to experience replay, where $\textit{replay memory size} = 10000$ defines the number of experiences the agent is capable of storing, and $\textit{replay batch size} = 1024$ defines the number of samples the agent uses at each time step to update the value function. 
The last are related to the epsilon-greedy exploration function, where we define an $\textit{epsilon initial} = 1$ as maximum epsilon value, an $\textit{epsilon final} = 0.01$ as epsilon minimum value, a percentage in which $\textit{epsilon decays} = 1\%$, and the interval of $\textit{time steps at each decay} = 128$.

We train our agent for the course of 64 thousand time steps in the environment.
Training statistics are gathered every 128 steps and are shown in Figure~\ref{fig:training-statistics}. 
Sub-figure~\ref{fig:rewards} shows the total reward accumulated by the agent at each 128 steps in the environment, which consistently improves over time and stabilizes after the 400th x-axis value. 
Sub-figure~\ref{fig:loss} shows the accumulated loss at each 128 steps in the environment, i.e. the errors in predictions of the value function during experience replay, and illustrates how it decreases towards zero as parameters are adjusted and the agent approximates the true value function.


To evaluate the agent behavior and the index configuration in which the agent is converging to, we plot in Figure~\ref{fig:training-indexes} each of the index configurations explored by the agent in the 64 thousand training steps.
Each index configuration is represented in terms of \textit{total indexes} and \textit{total optimal indexes} a configuration contains.
\textit{Total indexes} is simply a count on the number of indexes in the configuration, while \textit{total optimal indexes} is a count on the number of ground truth optimal indexes in the configuration.
The lines are smoothed using a running mean of the last 5 values, and a fixed red dashed line across the x-axis represents the configuration in which the agent should converge to.
As we can see, both the total amount of indexes \textit{and} the total optimal indexes converge towards the ground truth optimal indexes. 
That is, the agent learns both to keep the optimal indexes in the configuration, as well as to drop irrelevant indexes for the workload.

\subsection{\label{sec:performance-comparison}Performance Comparison}

\begin{figure}[b]
    \centering
    \begin{subfigure}[b]{0.23\textwidth}
        \centering
        \includegraphics[width=\textwidth]{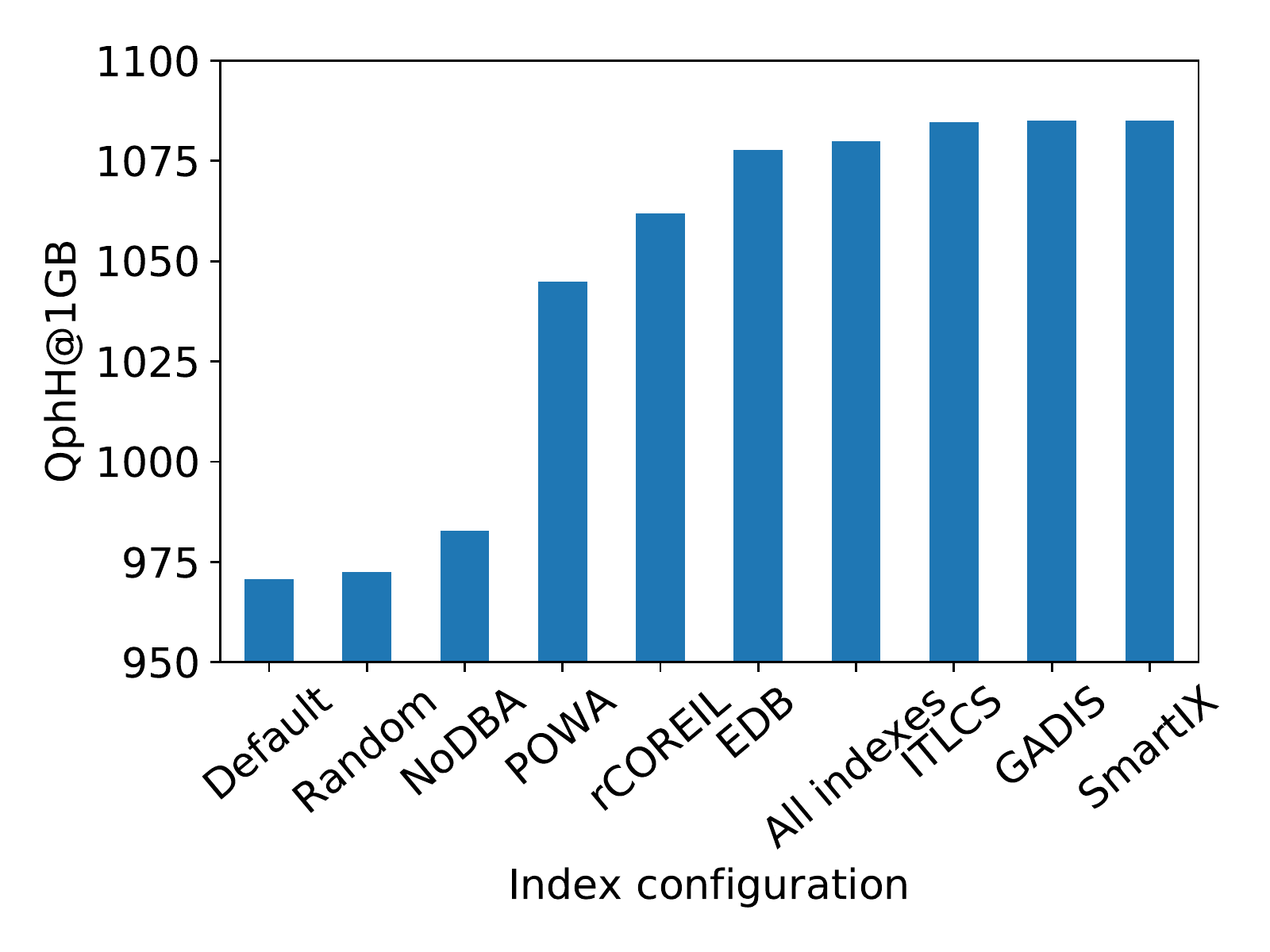}
        \vspace{-0.6cm}
        \caption{Query-per-hour metric}
        \label{fig:qphh}
    \end{subfigure}
    \begin{subfigure}[b]{0.23\textwidth}
        \centering
        \includegraphics[width=\textwidth]{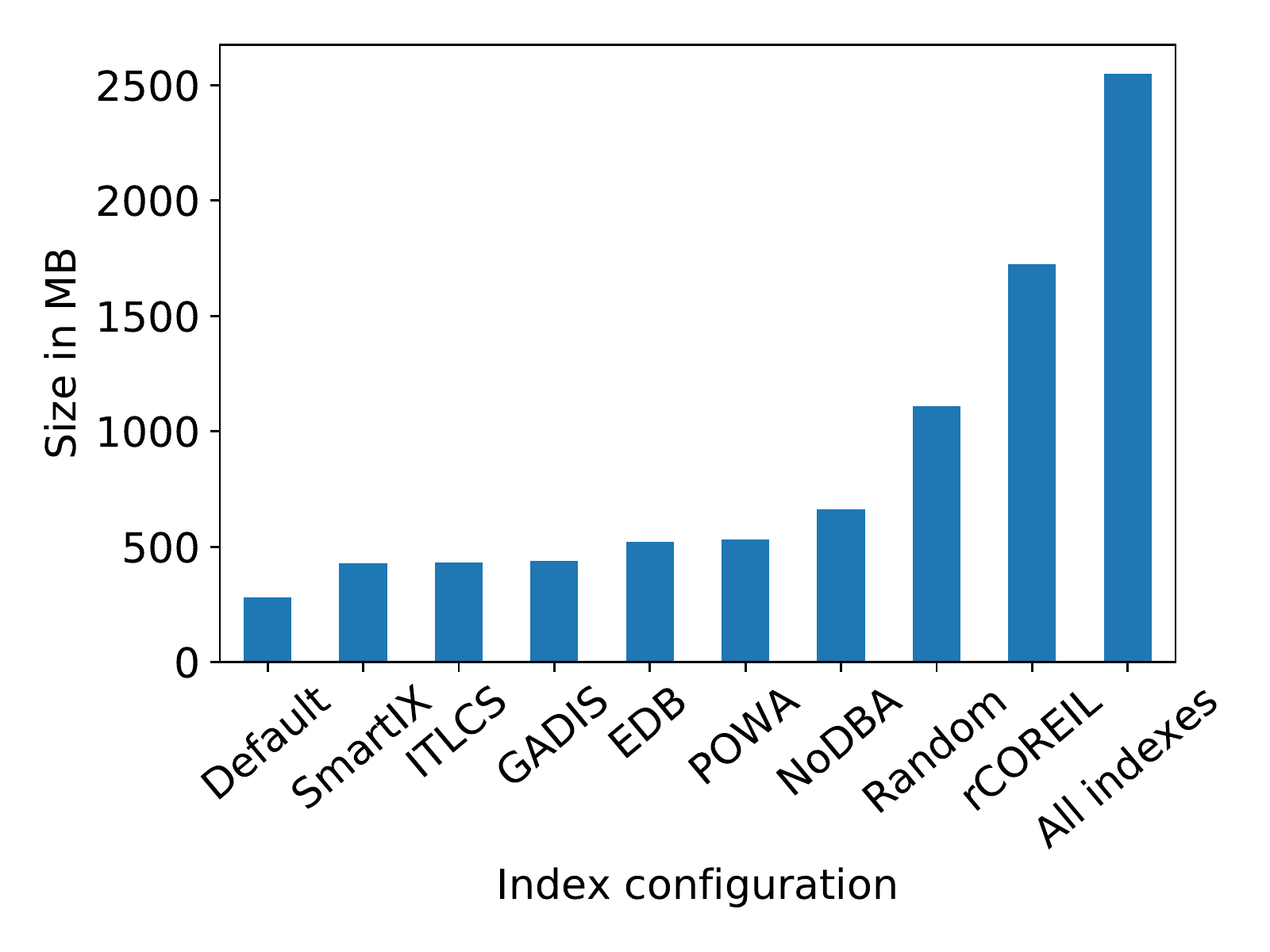}
        \vspace{-0.6cm}
        \caption{Index size (in MB)}
        \label{fig:size}
    \end{subfigure}
    \vspace{-0.1cm}
    \caption{Static index configurations results.}
    \label{fig:static-configurations}
\end{figure}

We now evaluate each baseline index configuration in comparison to the one in which our agent converged to in the last episode of training.
We show the TPC-H performance metric (QphH, i.e. the query-per-hour metric) and the index size of each configuration.
Figure~\ref{fig:qphh} shows the query-per-hour metric of each configuration (higher values denote better performance).
The plotted values are a result of a trimmed mean, where we run the TPC-H benchmark 12 times for each index configuration, removing the highest and the lowest result and averaging the 10 remaining results. 
Figure~\ref{fig:size} shows the disk space required for the indexes in each configuration (index size in MB), which allows us to analyze the trade-off in the number of indexes and the resources needed to maintain it.
In an ideal scenario, the index size is just the bare minimum to maintain the indexes that are necessary to support query performance.

Yet SmartIX achieves the best query-per-hour-metric, the two genetic algorithms \cite{neuhausgadis} and \cite{pedrozo2018adaptive} have both very similar query-per-hour and index size metrics in comparison to our agent. 
GADIS~\cite{neuhausgadis} itself uses a similar state-space model to SmartIX, with individuals being represented as binary vectors of the indexable columns.
The fitness function GADIS optimizes is the actual query-per-hour metric, and it runs the whole TPC-H benchmark every time it needs to compute the fitness function.
Therefore, it is expected that it finds an individual with a high performance metric, although it is unrealistic for real-world applications in production due to the computational cost of running the benchmark.

Indexing all columns is among the highest query-per-hour results and can seem to be a natural alternative to solve the indexing problem.
However, it results in the highest amount of disk used to maintain indexes stored. 
Such alternative is less efficient in a query-per-hour metric as the benchmark not only takes into account the performance of \textsc{select} queries, but also \textsc{insert} and \textsc{DELETE} operations, whose performance is affected by the presence of indexes due to the overhead of updating and maintaining the structure when records change~\cite[Ch. 8, p. 290-291]{ramakrishnan2000database}.
It has the lowest ratio value due to the storage it needs to maintain indexes.

While rCOREIL~\cite{basu2016regularized} is the most competitive reinforcement learning method in comparison to SmartIX, the amount of storage used to maintain its indexes is the highest among all baselines (except for having all columns indexed). 
rCOREIL does not handle whether primary and foreign key indexes are already created, causing it to create duplicate indexes.
The policy iteration algorithm used in rCOREIL is a dynamic programming method used in reinforcement learning, which is characterized by complete sweeps in the state space at each iteration in order to update the value function.
Since dynamic programming methods are not suitable to large state spaces~\cite[Ch. 4, p. 87]{sutton2018reinforcement}, this can become a problem in databases that contain a larger number of columns to index.

Among the database advisors, the commercial tool EDB~\cite{EDB} achieves the highest query-per-hour metric in comparison to the open-source tool POWA~\cite{POWA}, while its indexes occupy virtually the same disk space.
Other baselines and related work are able to optimize the index configuration and have lightweight index sizes, but are not competitive in comparison to the previously discussed methods in terms of the query-per-hour performance metric.
Finally, among all the baselines, the index configuration obtained using SmartIX not only yields the best query-per-hour metric but also the smallest index size (except for the default configuration), i.e. it finds the balance between performance and storage, as shown in the ratio plot.

\section{\label{sec:dynamic-configs}Dynamic configurations}

This section aims to evaluate the behavior of algorithms that generate policies, i.e. generate a function that guides an agent's behavior.
The three algorithms that generate policies are SmartIX, rCOREIL, and NoDBA. 
The three are reinforcement learning algorithms, although using different strategies (see Sec.~\ref{sec:related-work}). 
While rCOREIL and SmartIX show a more interesting and dynamic behavior, the NoDBA algorithm shows a fixed behavior and keeps only three columns indexed over the whole time horizon, without changing the index configuration over time (see its limitations in Sec.~\ref{sec:related-work}).
Therefore, we do not include NoDBA in the following analysis and focus the discussion on rCOREIL and SmartIX.

\subsection{Fixed workload}

\begin{figure}[tb]
    \centering
    \begin{subfigure}[b]{0.23\textwidth}
        \centering
        \includegraphics[width=\textwidth]{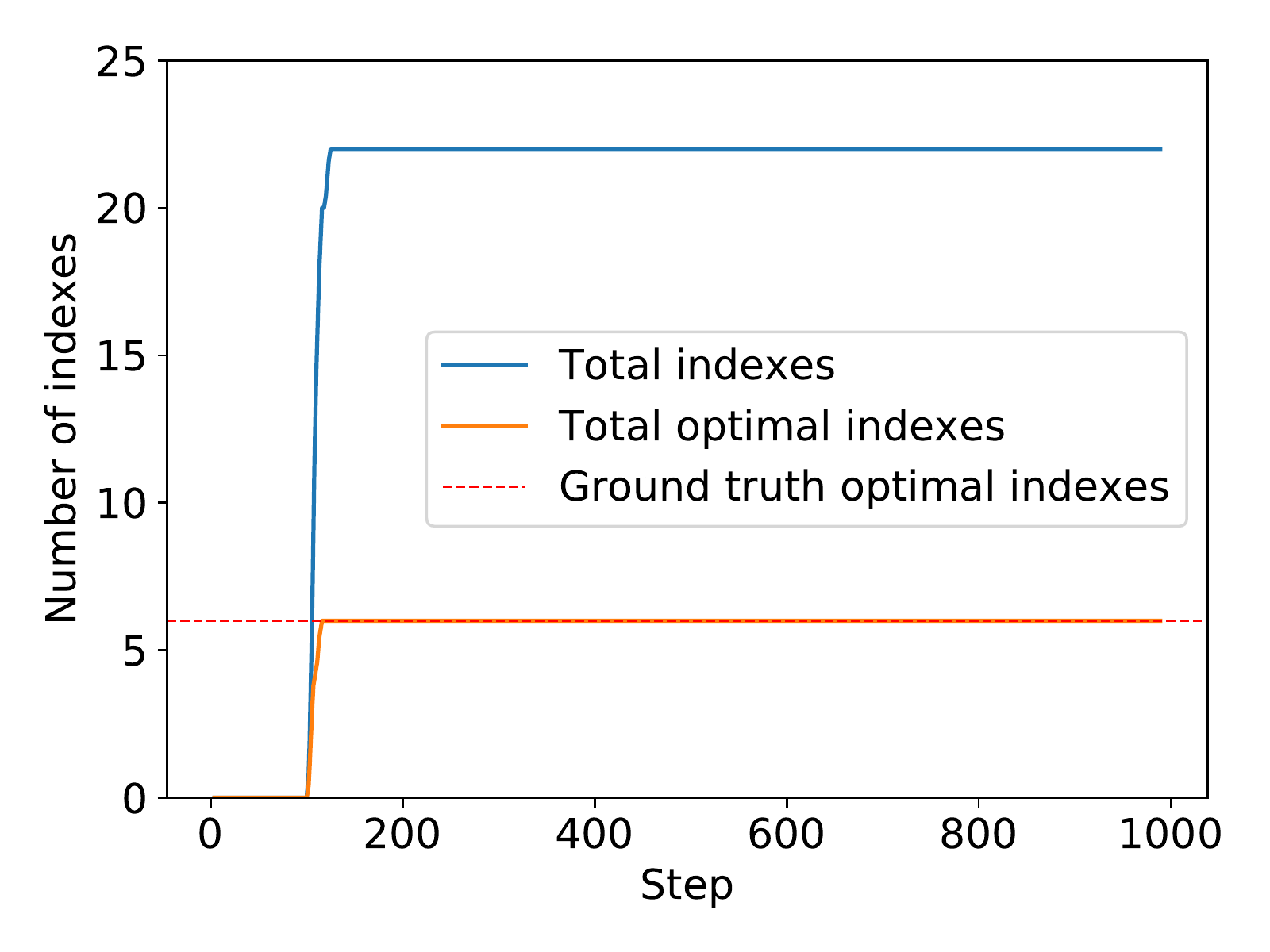}
        \vspace{-0.6cm}
        \caption{rCOREIL}
        \label{fig:fixed-workload-rcoreil}
    \end{subfigure}
    \begin{subfigure}[b]{0.23\textwidth}
        \centering
        \includegraphics[width=\textwidth]{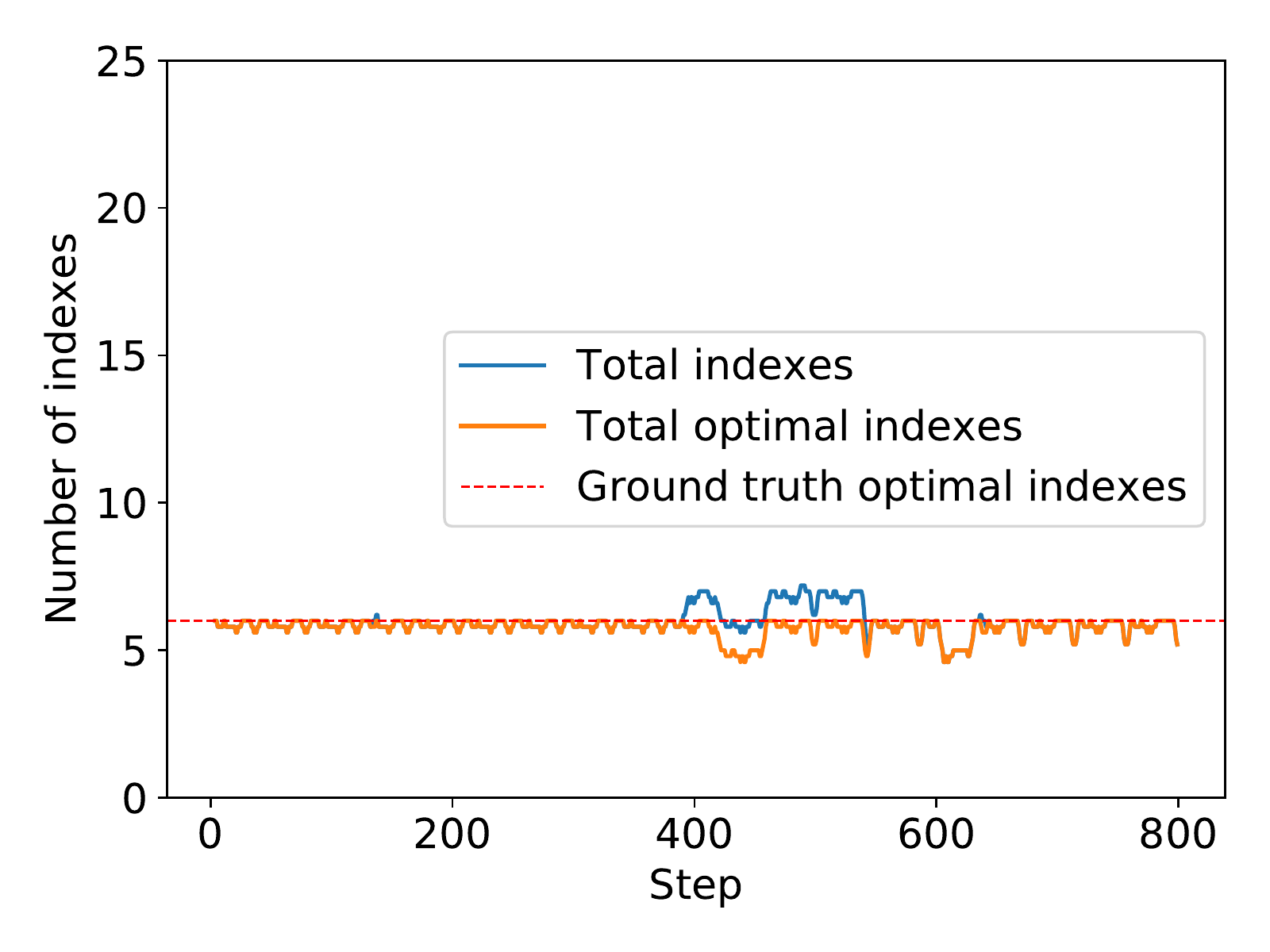}
        \vspace{-0.6cm}
        \caption{SmartIX}
        \label{fig:fixed-workload-smartix}
    \end{subfigure}
    \vspace{-0.1cm}
    \caption{Agent behavior with a fixed workload.}
    \label{fig:fixed-workload}
\end{figure}

We now evaluate the index configuration of rCOREIL and SmartIX over time while the database receives a fixed workload of queries. 
Figure~\ref{fig:fixed-workload} shows the behavior of rCOREIL and SmartIX, respectively. 
Notice that rCOREIL takes some time to create the first indexes in the database, after receiving about 150 queries, while SmartIX creates indexes at the very beginning of the workload.
On the one hand, rCOREIL shows a fixed behavior maintains all ground truth optimal indexes, but it creates a total of 22 indexes, 16 of those being unnecessary indexes and the remaining 6 are optimal indexes.
On the other hand, SmartIX shows a dynamic behavior and consistently maintains 5 out of the 6 ground truth optimal indexes, and it does not maintain unnecessary indexes throughout most of the received workload. 

\subsection{Shifting workload}

We now evaluate the algorithm's behavior while receiving a workload that shifts over time. 
To do so, we divide the 22 TPC-H queries into two sets of 11 queries, where for each set there is a different ground truth set of indexes.
That is, out of the 6 ground truth indexes from the previous fixed workload, we now separate the workload to have 3 indexes that are optimal first set of queries, and 3 other indexes that are optimal for the second set of queries.
Therefore, we aim to evaluate whether the algorithms can adapt the index configuration over time when the workload shifts and a different set of indexes is needed according to each of the workloads.

\begin{figure}[tb]
    \centering
    \begin{subfigure}[b]{0.23\textwidth}
        \centering
        \includegraphics[width=\textwidth]{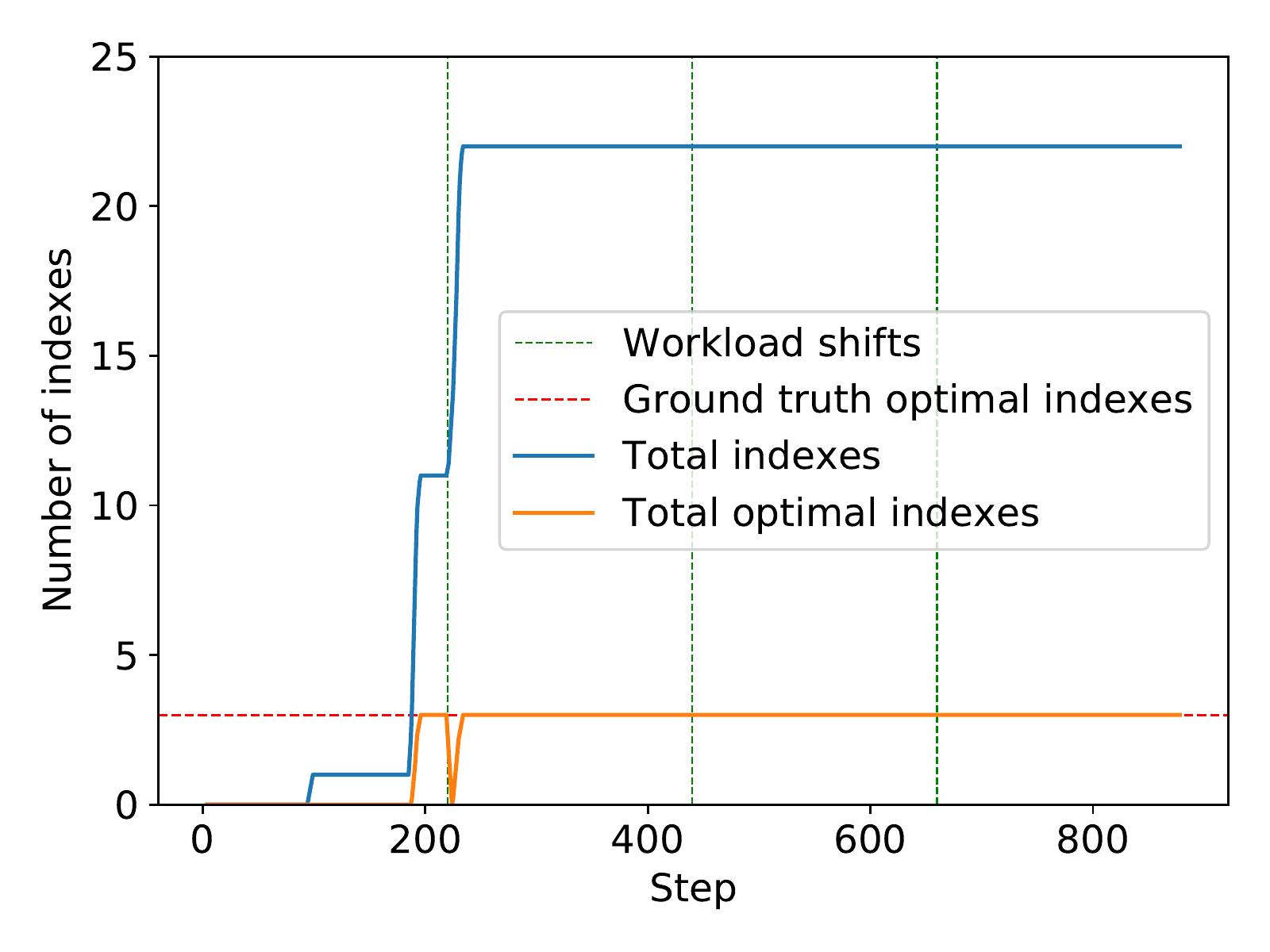}
        \vspace{-0.6cm}
        \caption{rCOREIL}
        \label{fig:shifting-workload-rcoreil}
    \end{subfigure}
    \begin{subfigure}[b]{0.23\textwidth}
        \centering
        \includegraphics[width=\textwidth]{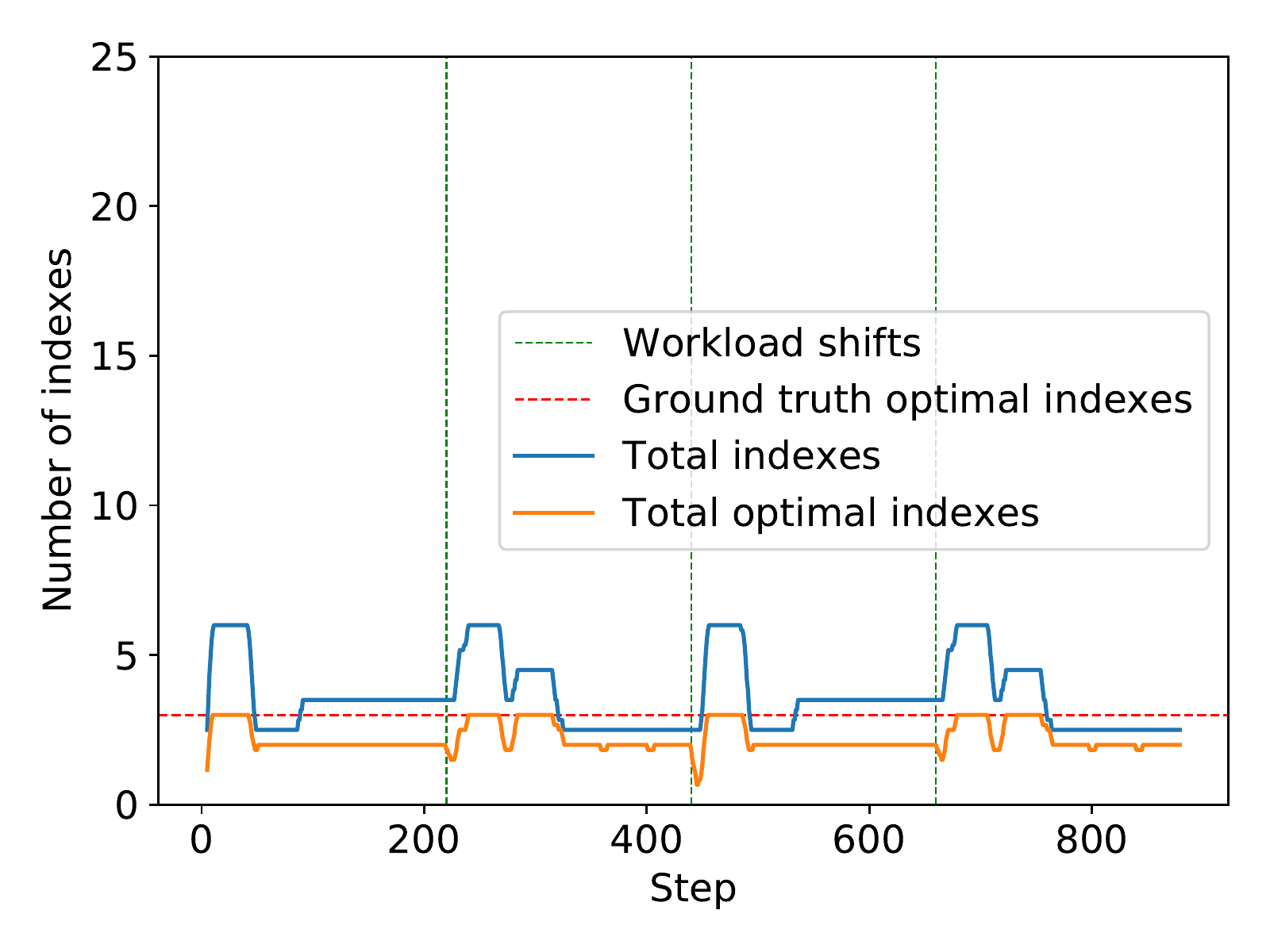}
        \vspace{-0.6cm}
        \caption{SmartIX}
        \label{fig:shifting-workload-smartix}
    \end{subfigure}
    \vspace{-0.1cm}
    \caption{Agent behavior with a shifting workload.}
    \label{fig:shifting-workload}
\end{figure}

The behavior of each algorithm is shown in Figure~\ref{fig:shifting-workload}. 
The vertical dashed lines placed along the x-axis represent the time step where the workload shifts from one set of queries to another, and therefore the set of ground truth optimal indexes also changes.
On the one hand, notice that rCOREIL shows a similar behavior from the one in the previous fixed workload experiment, in which it takes some time to create the first indexes, and then maintains a fixed index configuration, not adapting as the workload shifts. 
On the other hand, SmartIX shows a more dynamic behavior with regard to the shifts in the workload. 
Notice that, at the beginning of each set of queries in the workload, there is a peak in the total indexes, which decreases as soon as the index configuration adapts to the new workload and SmartIX drops the unnecessary indexes with regard to the current workload. 
Even though rCOREIL maintains all 3 ground truth indexes over time, it still maintains 16 unnecessary indexes, while SmartIX consistently maintains 2 out of 3 ground truth optimal indexes and adapts as the workload shifts.

\subsection{\label{sec:scaling-up}Scaling up database size}

\begin{figure}
    \centering
    \begin{subfigure}[b]{0.23\textwidth}
        \centering
        \includegraphics[width=\textwidth]{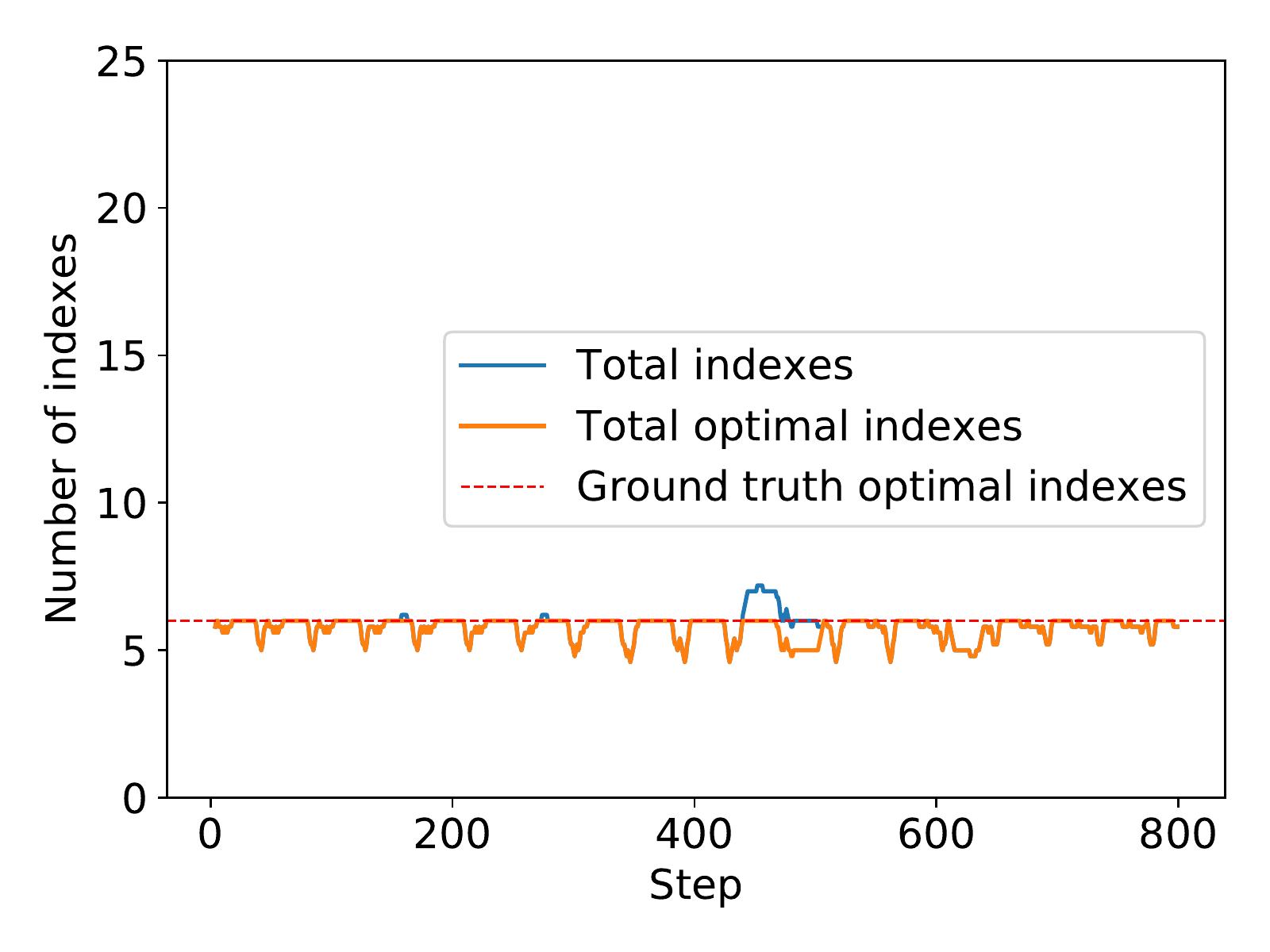}
        \vspace{-0.6cm}
        \caption{10GB TPC-H database.}
        \label{fig:fixed-workload-smartix-10gb}
    \end{subfigure}
    \begin{subfigure}[b]{0.23\textwidth}
        \centering
        \includegraphics[width=\textwidth]{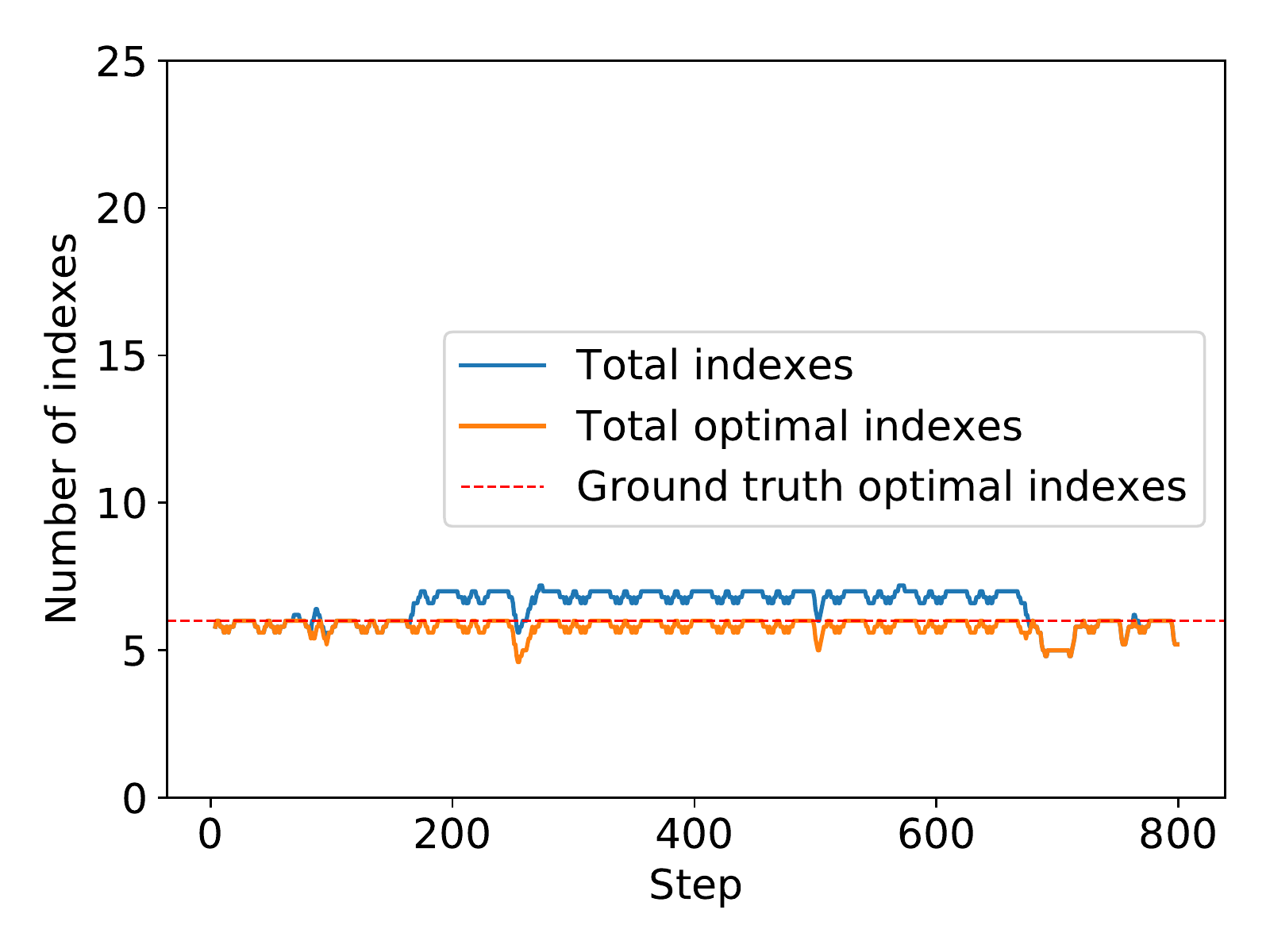}
        \vspace{-0.6cm}
        \caption{100GB TPC-H database.}
        \label{fig:fixed-workload-smartix-100gb}
    \end{subfigure}
    \vspace{-0.1cm}
    \caption{Agent behavior in larger databases.}
    \label{fig:larger-databases}
\end{figure}

In the previous sections, we showed that the SmartIX architecture can consistently achieve near-optimal index configurations in a database of size 1GB. 
In this section, we report experiments on indexing larger databases, where we transfer the policy trained in the 1GB database to perform indexing in databases with size 10GB and 100GB.
We plot the behavior of our agent in Figure~\ref{fig:larger-databases}.

As we can see, the agent shows a similar behavior to the one using a 1GB database size reported in previous experiments.
The reason is that both the state features and the reward function are not influenced by the database size.
The only information relevant to the state and the reward function is the current index configuration and the workload being received.
Therefore, we can successfully transfer the value function learned in smaller databases to index larger databases, consuming fewer resources to train the agent.

\section{\label{sec:related-work}Related Work}

Machine learning techniques are used in a variety of tasks related to database management systems and automated database administration~\cite{van2017automatic}.
One example is the work from Kraska et al.~\cite{kraska2018case}, which outperforms traditional index structures used in current DBMS by replacing them with learned index models, having significant advantages under particular assumptions.
Pavlo et. al~\cite{pavlo2017self} developed Peloton, which autonomously optimizes the system for incoming workloads and uses predictions to prepare the system for future workloads.
In this section, though, we further discuss related work that focused on developing methods for optimizing queries through automatic index tuning. 
Specifically, we focus our analysis on work that based their approach on reinforcement learning techniques.

Basu et al.~\cite{basu2016regularized} developed a technique for index tuning based on a cost model that is learned with reinforcement learning. 
However, once the cost model is known, it becomes trivial to find the configuration that minimizes the cost through dynamic programming, such as the policy iteration method used by the authors. 
They use DBTune~\cite{DBTune} to reduce the state space by considering only indexes that are recommended by the DBMS. 
Our approach, on the other hand, focuses on finding the optimal index configuration without having complete knowledge of the environment and without heuristics of the DBMS to reduce the state space. 

Sharma et al.~\cite{sharma2018case} use a cross-entropy deep reinforcement learning method to administer databases automatically.
Their set of actions, however, only include the creation of indexes, and a budget of 3 indexes is set to deal with space constraints and index maintenance costs.
Indexes are only dropped once an episode is finished.
A strong limitation in their evaluation process is to only use the \textsc{Lineitem} table to query, which does not exploit how indexes on other tables can optimize the database performance, and consequently reduces the state space of the problem.
Furthermore, they do not use the TPC-H benchmark performance measure to evaluate performance but use query execution time in milliseconds.


Other papers show that reinforcement learning can also be explored in the context of query optimization by predicting query plans: 
Marcus et al.~\cite{marcus2018deep} proposed a proof-of-concept to determine the join ordering for a fixed database; 
Ortiz et al.~\cite{ortiz2018learning} developed a learning state representation to predict the cardinality of a query. 
These approaches could possibly be used alongside ours, generating better plans to query execution while we focus on maintaining indexes that these queries can benefit from.

\section{\label{sec:conclusion}Conclusion}

In this research, we developed the SmartIX architecture for automated database indexing using reinforcement learning. 
The experimental results show that our agent consistently outperforms the baseline index configurations and related work on genetic algorithms and reinforcement learning.
Our agent is able to find the trade-off concerning the disk space its index configuration occupies and the performance metric it achieves.
The state representation and the reward function allows us to successfully index larger databases while training in smaller databases and consuming fewer resources. 

Regarding the limitations of our architecture, we do not yet deal with composite indexes due to the resulting state space of all possible indexes that use two or more columns.
Our experiments show results using workloads that are read-intensive (i.e. intensively fetching data from the database), which is exactly the type of workload that benefits from indexes.
However, experiments using write-heavy workloads (i.e. intensively writing data to the database) can be interesting to verify whether the agent learns to avoid indexes in write-intensive tables.
Considering these limitations, in future work, we plan to: (1) investigate techniques that allow us to deal with composite indexes; (2) improve the reward function to provide feedback in case of write-intensive workloads; (3) investigate pattern recognition techniques to predict incoming queries to index ahead of time; and (4) evaluate SmartIX on big data ecosystems (e.g. Hadoop).

Finally, our contributions include: (1) a formalization of a reward function shaped for the database indexing problem, independent of DBMS's statistics, that allows the agent to adapt the index configuration according to the workload; (2) an environment representation for database indexing that is independent of schema or DBMS; and (3) a reinforcement learning agent that efficiently scales to large databases, while trained in small databases consuming fewer resources. 
At last, as a result of this research, we published a paper at the Applied Intelligence journal~\cite{licks2020smartix}.

In closing, we envision this kind of architecture being deployed in cloud platforms such as Heroku and similar platforms that often provide database infrastructure for various clients' applications. 
The reality is that these clients do not prioritize, or it is not in their scope of interest to focus on database management. 
Especially in the case of early-stage start-ups, the aim to shorten time-to-market and quickly ship code motivates externalizing complexity on third party solutions~\cite{giardino2016software}. 
From an overall platform performance point of view, having efficient database management results in an optimized use of hardware and software resources.
Thus, in the lack of a database administrator, the SmartIX architecture is a potential stand-in solution, as experiments show that it provides at least equivalent and often superior indexing choices compared to baseline indexing recommendations.

\bibliographystyle{aaai}
\bibliography{main}

\end{document}